\documentclass[proof]{pasj01}
\usepackage[utf8]{inputenc}
\usepackage{graphicx}
\usepackage{multirow}
\usepackage{xcolor}
\usepackage{caption}
\usepackage{comment}
\usepackage[margin=22mm]{geometry} 
\usepackage{lineno}

\begin{document}
\title{Detection of the longest periodic variability in 6.7 GHz methanol masers G5.900$-$0.430}

\author{Yoshihiro \textsc{tanabe}\altaffilmark{1}}
\author{Yoshinori \textsc{yonekura}\altaffilmark{1}}
\author{MacLeod \textsc{c. gordon}\altaffilmark{2, 3}}

\altaffiltext{1}
{Center for Astronomy, Ibaraki University, 2-1-1 Bunkyo, Mito, Ibaraki 310-8512, Japan }
\altaffiltext{2}
{The Open University of Tanzania, P.O. Box 23409, Dar-Es-Salaam, Tanzania}
\altaffiltext{3}
{SARAO, Hartebeesthoek Radio Astronomy Observatory, PO Box 443, Krugersdorp, 1741, South Africa}

\email{yoshihiro.tanabe.ap@vc.ibaraki.ac.jp}

\maketitle

\begin{abstract}
Long term monitoring observations with the Hitachi 32-m radio telescope of the 6.7 GHz methanol masers associated with the high mass star-forming region G5.900$-$0.430 is presented. 
A period of flux variability at approximately 1260 days, is detected in the features at $V_{\rm LSR}$ = 9.77 and 10.84 km s$^{-1}$ while a secondary shorter period, 130.6 days, is determined for the 0.66 km s$^{-1}$ feature.
This is only the second source which has two different periods.
The period of $\sim$1260 days is approximately twice as long as the longest known period of 6.7 GHz methanol masers.
The variability pattern of the symmetric sine curves and the consistency with the expected period-luminosity relation suggest that the mechanism of maser flux variability of 9.77 and 10.84  km s$^{-1}$ features in this source can be explained by protostellar pulsation instability.
On the other hand, because the 0.66 km s$^{-1}$ feature has an intermittent and asymmetric variability profile, we propose this feature is explained by the CWB or spiral shock models.
Obtaining the spatial distribution of the 0.66 km s$^{-1}$ feature using an VLBI will lead to a better understanding of this source.

\end{abstract}
%\linenumbers

\section{Introduction}
High-mass stars have significant impact on their surrounding environment through various feedback mechanisms, e.g., stellar winds, ultraviolet radiation, and supernovae.
However, our understanding of the formation processes of high mass stars remains inadequate, hampered by the observational difficulties of their birth in distant, very deeply embedded dense gas.
The Class II 6.7 GHz methanol maser is a well-known tracer of high-mass star-forming regions (HMSFRs) (e.g, Menten\ 1991; Minier et al.\ 2003; Pandian et al.\ 2007; Breen et al.\ 2013).
Emission arises in most cases from regions of an interface between the protostellar disk and envelope at a typical distance of $\sim$1000 au from a protostar (Bartkiewicz et al.\ 2016).
The 6.7 GHz maser is sensitive to the local physical conditions around high-mass protostars (e.g., Cragg,  Sobolev, and Godfrey\ 2005; Sugiyama et al.\ 2008; Moscadelli et al.\ 2017; Burns et al.\ 2020), and thus is an excellent observational probe to study the high-mass star formation process.

Goedhart et al.\ (2003) first discovered a periodic flux variation of 243.3 days in the 6.7 GHz methanol masers associated with HMSFR G9.62$+$0.20E.
Another 27 periodic methanol maser sources have been reported (Goedhart et al.\ 2004, 2009, 2014; Szymczak et al.\ 2011, 2014, 2015, 2016; Araya et al.\ 2010; Fujisawa et al.\ 2014; Maswanganye et al.\ 2015, 2016; Sugiyama et al.\ 2015, 2017; Proven-Adzri et al.\ 2019; Olech et al.\ 2019, 2022), and their periods range from 23.9 to 668 days.
Systematic monitoring observations toward large sample have contributed significantly to the discovery of maser periodicity (e.g., Goedhard et al.\ 2004; Sugiyama et al.\ 2018; Szymczak et el.\ 2018; Olech et al.\ 2022).
The variation pattern in the time series of the 6.7 GHz methanol maser are classified into two types: continuous (e.g., G188.95$+$0.89 reported by Goedhard et al.\ 2014) and intermittent (e.g., IRAS 22198$+$6336$=$G107.298$+$0.5639 reported by Fujisawa et al.\ 2014).
Such differences in variation patterns are thought to be due to differences in the origin of 6.7 GHz maser periodicity.
The mechanism of these maser periodicity remains an open question, and several explanations have been proposed;
colliding wind binary (CWB) system (van der Walt\ 2011; van der Walt et al.\ 2016), protostellar pulsation (Inayoshi et al.\ 2013), spiral shock (Parfenov \& Sobolev\ 2014), periodic accretion in a circumbinary system (Araya et al.\ 2010), or a very young low-mass companion blocking the ultraviolet radiation from the high-mass star in an eclipsing binary (Maswanganye et al.\ 2015).
These models can explain the maser fluctuations of some sources well, however, there is no clear consensus.

In this paper, we present the new discovery of periodicity in 6.7 GHz methanol maser source in HMSFR G5.900$-$0.430, and discuss the mechanism of its flux variability.

\section{Observation}
Monitoring observations of the 6.7 GHz methanol maser were made with the Hitachi 32m telescope of Ibaraki station, a branch of the Mizusawa VLBI Observatory of the National Astronomical Observatories Japan (NAOJ), operated jointly by Ibaraki University and NAOJ (Yonekura et al. 2016).
This is as a part of the Ibaraki 6.7 GHz Methanol Maser Monitor\ (iMet) program\footnote{http://vlbi.sci.ibaraki.ac.jp/iMet/}.

Monitoring observations of G5.900$-$0.430 began on 2013 Jan.\ 03 (modified julian day (MJD) $=$ 56295).
The cadence of observations is once par every $\sim$10 days from the start of the monitoring observations to 2015 Aug.\ 08 (MJD $=$ 57242), and once par every $\sim$5 days from 2015 Sep.\ 19 (MJD $=$ 57284) to the present. 
Observations after 2014 May 08 (MJD = 56785) were made at about the same azimuth ($\sim$187$\degree$) and elevation angle ($\sim$27$\degree$) to minimise intensity variations due to systematic telescope pointing errors.
The half-power beam width of the telescope is $\sim$ \timeform{4.6'} with the pointing accuracy better than $\sim$ \timeform{30"}. 
The coordinates of G5.900$-$0.430 adopted in observations are ${\rm R.A.} =$ \timeform{18h00m 40s.86}, Dec. $=$ \timeform{-24D04'20.8"}  (J2000.0)\ (Caswell\ 2009; Caswell et al.\ 2010).

Observations are made by using a position-switching method. The OFF position is set to $\Delta {\rm R.A.} = +\ \timeform{60'}$ from the target source. The integration time per observation is 5 minutes for both the ON and OFF positions.
A single left circular polarization (LCP) signal was sampled at 64 Mbps (16 mega-samples per second with 4 bit sampling) by using a K5/VSSP32 sampler.
The recorded bandwidth is 8 MHz (RF:6664-6672 MHz) and they are divided into 8192 channels.
After averaging over 3 channels, the 1-sigma root-mean-squares noise level is approximately 0.3 Jy and the velocity resolution is 0.13 km s$^{-1}$.
The antenna temperature was measured by the chopper-wheel method and the system noise temperature toward the zenith after the correction for the atmosphere opacity (${T}^{*}_{\rm sys}$) is typically 25--35 K.
In our monitoring program, we observe $\sim$60 methanol maser sources per day, of which the variation of the flux density of sources that do not show the intrinsic variation are less than $\sim$20\%.

\section{Result}
\begin{figure}[h!tb]
 \begin{center}
  \includegraphics[width=8.5cm, bb= 0 0 576 432]{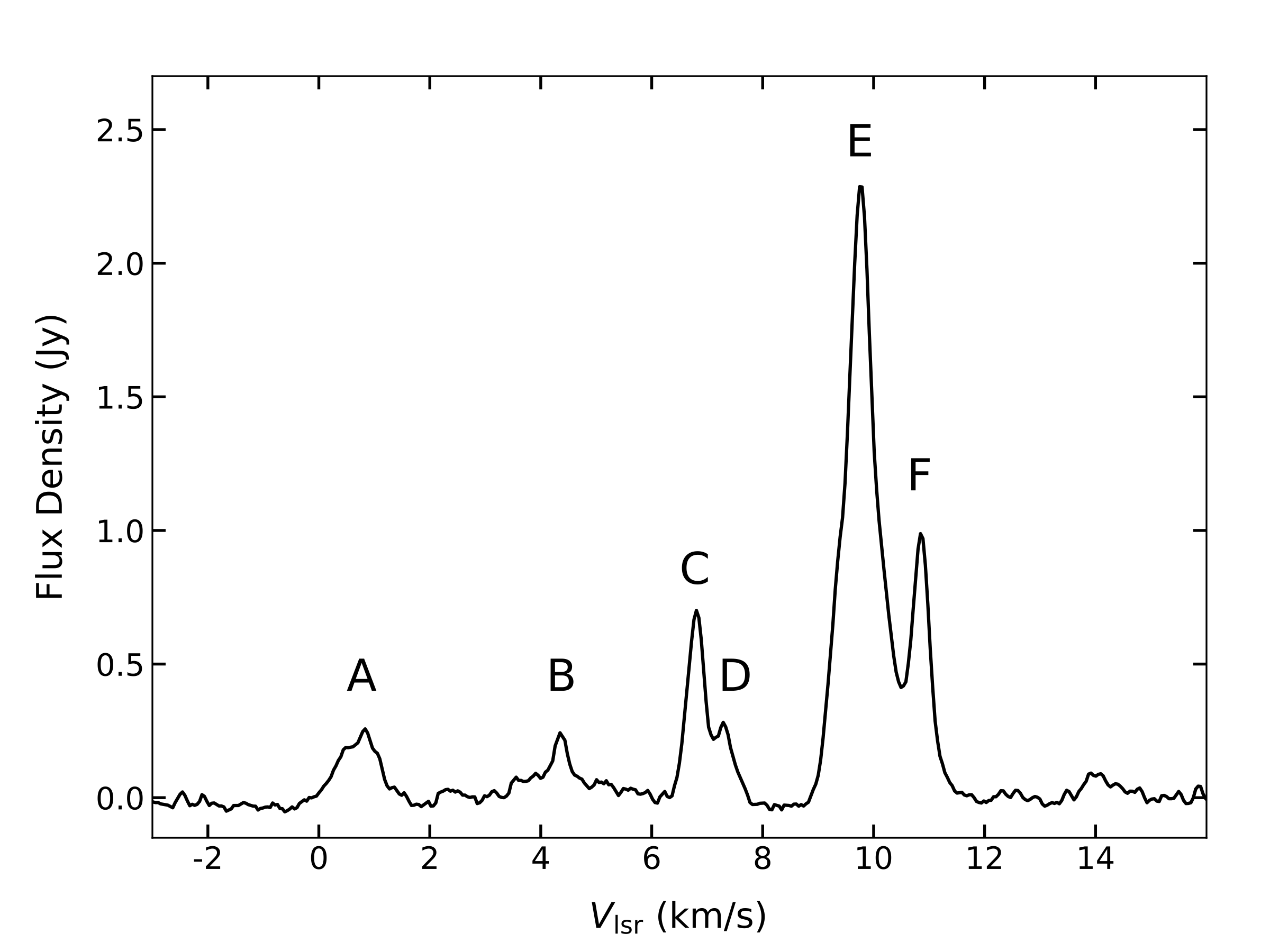}
   \caption{Averaged spectrum of the 6.7 GHz methanol maser associated with G5.900$-$0.430.
   All 437 scans of 5 minutes duration are averaged and 3$\sigma$ detection limit is 0.042 Jy.
   Labels A, B, C, D, E and F indicate spectral components at $V_{\rm LSR} =$ 0.66, 4.35, 6.76, 7.29, 9.77, and 10.84 km s$^{-1}$
   , respectively.}
    \label{sp}
 \end{center}
\end{figure}

\begin{figure*}[h!tb]
 \begin{center}
 \includegraphics[width=18cm, bb=0 0 720 405]{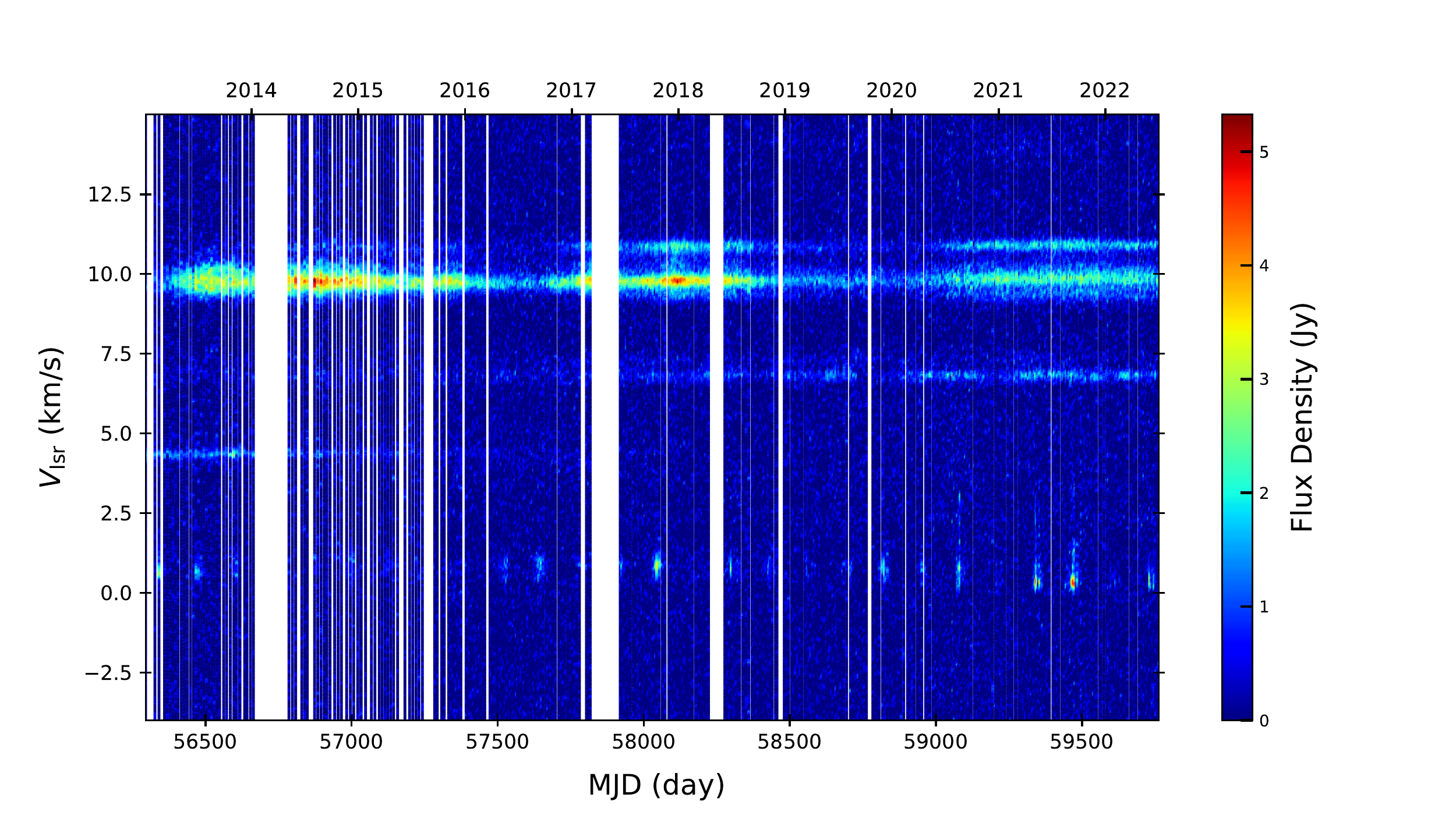}
   \caption{Dynamic spectrum compiling all 437 scans of the 6.7 GHz methanol masers associated with G5.900-0.430.
   Labels on the upper axis indicate beginning of the each year.
   White blanks indicate no observations were made.}
    \label{dy}
 \end{center}
\end{figure*}

The averaged spectrum and dynamic spectrum compiling all 437 scans of G5.900-0.430 is presented in Figures \ref{sp} and \ref{dy}. 
In this source, seven velocity features are distributed from $V_{\rm LSR}$ $\sim$0 km s$^{-1}$ to $\sim$14
km s$^{-1}$. 
According to Caswell et al.\ (2010), the 6.7 km s$^{-1}$ feature (C and D in Figure \ref{sp}) is associated with HMSFR G5.885$-$0.393, approximately 2.3 arcmin apart from G5.900$-$0.430.
Therefore we exclude these features from the discussion below.

\begin{figure*}[h!tb]
 \begin{center}
  \includegraphics[width=18cm, bb=0 0 1296 720]{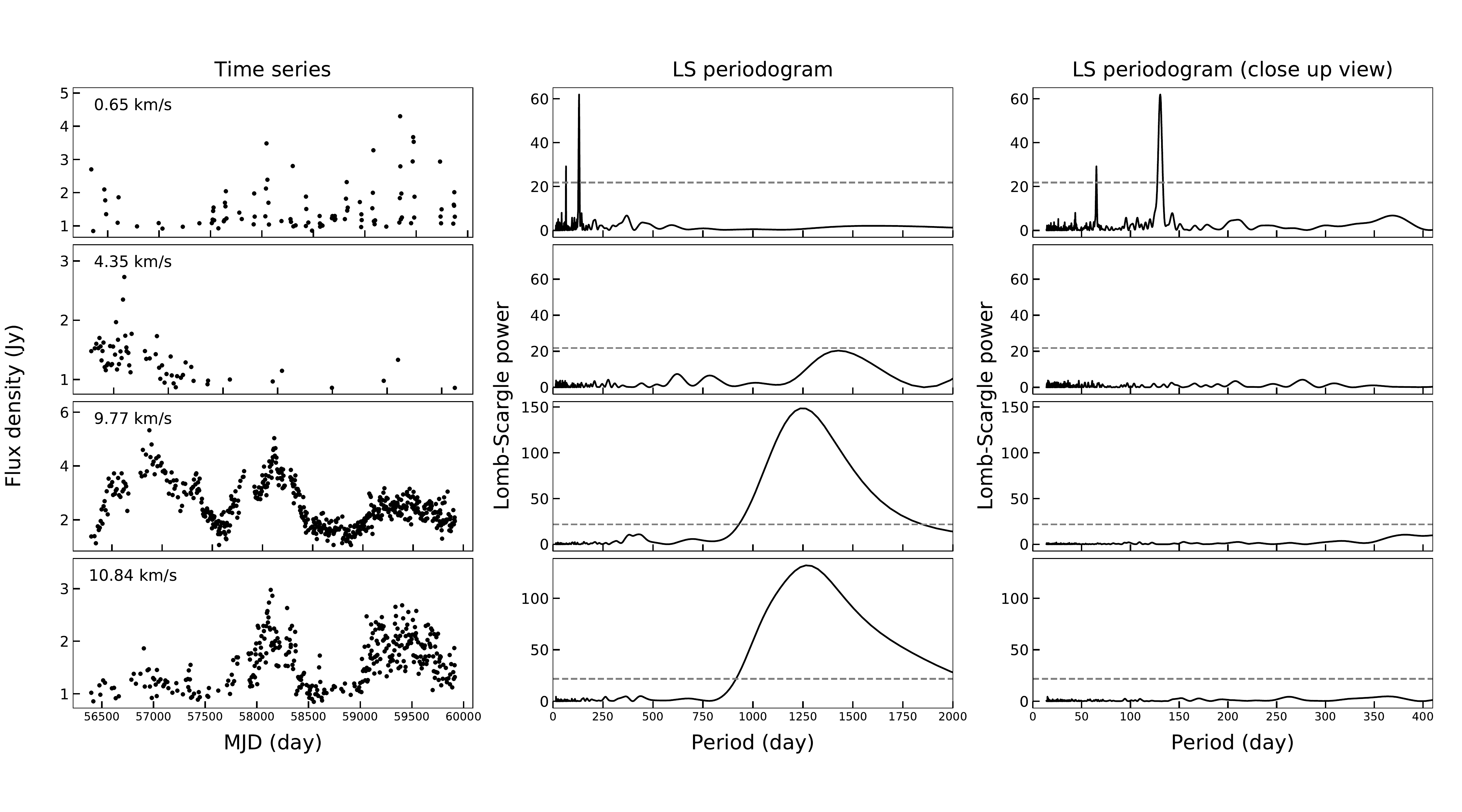}
   \caption{Time series and results of LS analysis for all velocity components in G5.900$-$0.340.  
   Left column show the time series of flux density of each velocity feature.
   We excluded the data points whose flux densities are less than 3 $\sigma$.
   The second and third column show the Lomb-Scargle power spectra plots.
   The dotted lines in second and third columns represent the 0.01\% false alarm probability (probability of judging noise as a real signal) levels and if the peak value of the power spectrum is higher than the dotted line, the obtained period is reliable.}
    \label{sr}
 \end{center}
\end{figure*}

\begin{figure*}[h!tb]
 \begin{center}
  \includegraphics[width=18cm, bb=0 0 1296 288]{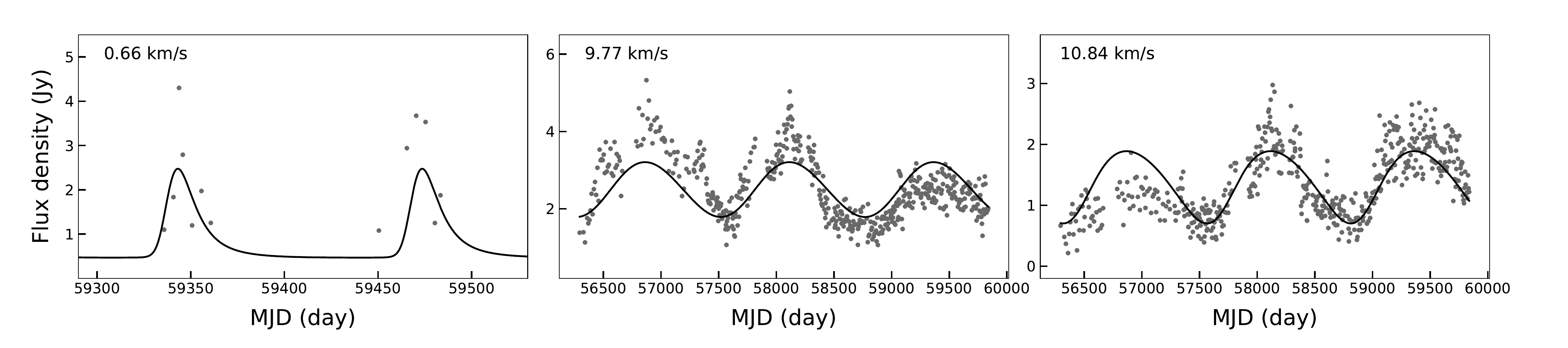}
   \caption{Results of power function fitting.
   Gray points represent time series of each periodic component excluding the data points whose flux densities are less than 3 $\sigma$ and black lines show the best fitting of periodic power function.
   It should be noted that, the entire data were used for the fitting, but for 0.66 km s$^{-1}$ feature (left panel), we show here only the data from MJD = 59290 to MJD = 59530.}
    \label{fit}
 \end{center}
\end{figure*}

The results of periodicity analysis are summarized in Figures \ref{sr}, \ref{fit} and Table \ref{tabl}.
The periodicity was estimated eby mploying the Lomb-Scargle (LS) periodogram method (Lomb\ 1976 and Scargle\ 1982) and  the asymmetric power function given by the equation adopted from Szymczak et al. (2011):
\begin{equation}
S(t)=C\times {\rm exp}^{s(t)} +D ,
\end{equation}
where {\it C} and {\it D} are constants and 
\begin{equation}
s(t)=\frac{-B\ {\rm cos}(2\pi\frac{t}{P}+\phi)}{1-f\ {\rm sin}(2\pi\frac{t}{P} +\phi)}+A\ .
\end{equation}
Here {\it A} and {\it B} are constants, 
{\it P} is the period, $\phi$ is the phase at $t=0$, and $f$ is the asymmetric parameter defined as the rise time from the minimum to the maximum flux (Szymczak et al.\ 2011).
When $f = 0$, the power function is symmetric with respect to the peak.
The error of periods obtained by the LS method are estimated as the half width of half maximum (HWHM) of each peak in the periodgram.

\begin{figure*}[h!tb]
 \begin{center}
  \includegraphics[width=18cm, bb=0 0 720 324]{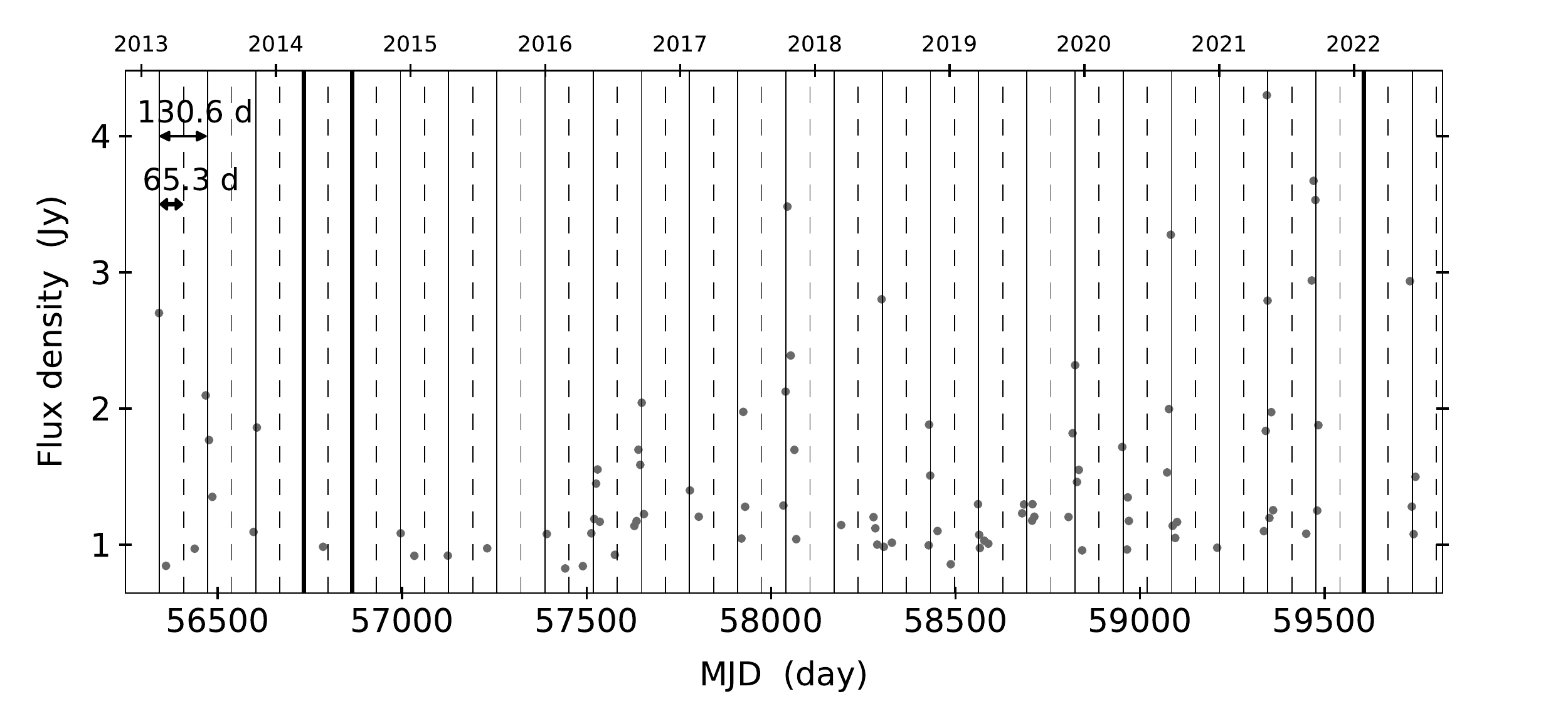}
   \caption{Time series of 0.66 km s$^{-1}$ feature.
   The solid lines show every 130.6 days from MJD $=$ 56342.3 which is the first peak of best-fit power function
   and the dotted lines is in the middle of the solid lines.
   Labels on the upper axis indicate
beginning of the each year.
The thick solid lines around MJD=56734, 56865, and 59607 indicate that the flux density over 3 sigma were not detected during the 8 days before and after the expected flare date.}
    \label{066}
 \end{center}
\end{figure*}

A periodic and intermittent profile is found at 0.66 km s$^{-1}$ feature (A in Figure \ref{sp}).
Periods of 130.6$\ \pm\ $2.8 and 65.2$\ \pm\ $0.7 days were determined from the LS periodicity analysis. Note the latter appears to be a harmonic of the former, i.e. $P=$130.6/n where n = 2.
In fact in figure \ref{066}, almost no flares can be detected in the period of 65.3 days (dashed line), which is half of 130.6 days (solid line).
Thus the 130.6 days is the most plausible period for 0.66 km s$^{-1}$ feature.
An asymmetric power function fitting determines a value of $P_{\rm fit}=$130.5 $\ \pm\ $ 0.1 days and $f=$0.73.
The periods obtained by the two methods are within the errors measured.
The value of $f$ suggests that the 0.66 km s$^{-1}$ feature experiences an asymmetrical temporal flux density variation; a rapid onset followed by a slower decline.
The HWHM of rising and decaying time obtained directly from fitting results are 8.16 and 13.88 days, respectively.

As shown in Figure \ref{066}, during our monitoring observation, there were 3 cases where no flares were detected at the expected flare date.
In the first two of these three cases, observations were not made within 5 days before and after the expected flare date due to maintenance or other reasons.
In the third case, we detected emissions with flux density of 2.5$\sigma$ and 2.6 $\sigma$ at 2 days before and 3 days after the expected flare date (MJD $=$ 59607), respectively, but no emissions above 3$\sigma$.

The 4.35 km s$^{-1}$ feature (B in Figure \ref{sp}) is generally below detection limits after MJD $=$ 57363, and no periodicity is detected for this feature.

The 9.77 km s$^{-1}$ and 10.84 km s$^{-1}$ features (E and F in Figure \ref{sp}) experience very long and continuous flux variation.
The estimated period from LS method and power function fitting for 9.77 km s$^{-1}$ feature are 1264.9$\ \pm\ $265.5 days and 1260.3 $\ \pm\ $8.6 days, respectively, and those for 10.84 km s$^{-1}$ feature are 1264.9$\ \pm\ $324.3 days and 1260.7$\ \pm\ $9.6 days, respectively.
The periods of each velocity features obtained by the two methods are consistent within the estimated errors.
The results of fitting show that the flux variation of these features are symmetric with $f=$ -0.20 and -0.11 for 9.77 and 10.84 km s$^{-1}$, respectively.

MacLeod et al.\ (2022) detected periodic flux variation of maser features with two different periods in HMSFR G9.62$+$0.20E.
According to MacLeod et al.\ (2022), 
G9.62$+$0.20E has secondary period of 52 days for 8.8 km s$^{-1}$ feature, along with the previously reported period of 243 days for another features.
G5.900$-$0.430 may be the second example which has a secondary period.
The temporal profiles of the 8.8 km s$^{-1}$ feature and the others are quite different, intermittent and continuous, respectively. 
This is similar to the present results for G5.900$-$0.430.
On the other hand, the relationship in the flux variation profiles and the periods shows a completely opposite trend.
In G5.900$-$0.430, the feature with a long period shows a continuous profile and the feature with a short period shows an intermittent profile, whereas in G9.62$+$0.20E has opposite tendency.

\begin{table}[h!tb]
{\scriptsize
\centering
\caption{Parameters of the periodicity}
  \begin{tabular}{ccccc}
  \hline
  $V_{\rm LSR}$&$ P_{\rm LS}$\footnotemark[$*$] &$P_{\rm fit}$\footnotemark[$\dag$]& {\it f}\footnotemark[$\ddag$]&$r^2$\footnotemark[$\S$] \\
  (km s$^{-1}$)&(days)&(days)&&\\\hline
  0.66 &   130.6\ (2.8) & 130.5\ (0.1) &  0.73\ (0.07) & 0.44\\
  9.77 & 1264.9\ (265.5)& 1260.3\ (8.6) & -0.20\ (0.07)   &0.67\\
  10.84 & 1264.9 (324.3)& 1260.7\ (9.6) & -0.11\ (0.07) &0.68
  \\\hline
  \end{tabular}
  \label{tabl}
  \begin{tabnote}
  \footnotemark[$*$]Period estimated by LS method.
  \footnotemark[$\dag$]Period estimated by power function fitting.
  \footnotemark[$\ddag$]Asymmetric parameter.
   \footnotemark[$\S$]Coefficient of determination.
  \end{tabnote}
  }
\end{table}

\section{Discussion}
\subsection{Spacial distribution of maser features}
The 6.7 GHz methanol maser emission of G5.900$-$0.430 was imaged by Hu et al.\ (2016) with the Karl G. Jansky Very Large Array (VLA).
The 0.66 km s$^{-1}$ feature was not detected and 4 km s$^{-1}$ to 10 km s$^{-1}$ features were imaged.
The observations by Hu et al.\ (2016) were made between MJD$=$55985 and 56033.
Assuming that the 0.66 km s$^{-1}$ feature has a period of 130 days before we started the monitoring,
the expected flare dates closest to dates of observations by Hu et al.\ (2016) are MJD$=$55952 and 56082, both of which are outside the observations of Hu et al.\ (2016).
Thus the observations of Hu et al.\ (2016) must have conducted in the quiet state of the  0.66 km s$^{-1}$ feature.
According to Hu et al.\ (2016), the  9-11 km s$^{-1}$ features have a special distribution of $\sim$ 600 mas, which correspond to $\sim$1700 au when the distance of 2.9 kpc is adopted
(see subsection \ref{pe}).
This is approximately comparable in size to typical maser distribution in high-mass protostar system.
Therefore, the high velocity features with a long period are thought to be excited by the central protostar.

The 4.35 km s$^{-1}$ feature is only just above our detection limits in our most recent single-dish observation.
These masers are superimposed on a bright, 7.4 Jy, H\emissiontype{II} region, obtained by two 1024 MHz sub-bands from 4.9840 to 6.0080 GHz and from 6.6245 to 7.6485 GHz, suggesting they are associated (Hu et al.\ 2016).

\subsection{Periodicity}\label{pe}
According to Olech et al.\ (2019 and 2022), among the known periodic maser sources, the longest period is 668 days in G196.45$-$1.68 (Goedhart et al.\ 2004) and only two sources have periods longer than 500 days. 
Our result increases the upper end of the range of known periods by a factor of two.
On the other hand, the 0.66 km s$^{-1}$ feature has a much shorter period of 130.6 days and the flaring is intermittent perhaps owing to the relative weakness of the feature.
The difference in the behavior of the variability of the low-velocity feature (0.66 km s$^{-1}$) and high-velocity features (9.77 and 10.84 km s$^{-1}$) suggests that the variability of the flux of these components may be the result of different mechanisms.
We discuss the mechanisms of maser variability below.

 The CWB model to cause periodic variations in HMSFRs was first proposed by var der Walt, Goedhart, and Gaylard\ (2009) and it was modeled more in detail by van der Walt\ (2011).
In this model, periodic wind interaction in a binary system generates changes in the background free-free emission which is amplified by the masers.
It requires a varying H\emissiontype{II} region, the free-free radiation acts as the seed photons. 
According to van der Walt et\ (2011), van der Walt et al.\ (2016) and Olech et al.\ (2022), 
the maser flare temporal profile may be explained by the CWB model where it is typified by a short onset and long decay.
This is the best explanation for the flare characteristics of the 0.66 km s$^{-1}$ feature in G5.900-0.430.
It is proposed that resulting variations of the associated background H\emissiontype{II} region cause the flaring in the 0.66 km s$^{-1}$ feature.
However, this model is unsuitable for the high-velocity features, which show continuous sinusoidal variation.

Inayoshi et al.\ (2013) presented that the high-mass protostars become pulsationally unstable under rapid mass accretion with rates of $\dot{M_{*}}\gtrsim 10^{-3}\ M_{\odot}\ {\rm yr}^{-1}$.
The protostar's luminosity varies periodically and as the temperature of the surrounding dust rises and falls to a temperature suitable for maser radiation, resulting in that the maser fluxes increase and decrease.
In this model, the flux variation of the maser is expected to be continuous, thus the pulsation model is the best describe the variation of the high-velocity features.
Inayoshi et al. (2013) also derived the period-luminosity relation \begin{equation}
{\rm log}\frac{L}{\ L_{\odot}} = 4.62 + 0.98\ {\rm log}\frac{P}{100 \ {\rm days}},
\end{equation}
where {\it L} is luminosity of the protostar and {\it P} is period expected from the maser.
For a period of 1260 days, the luminosity of G5.900$-$0.430 should be $\sim4.9\times$10$^{5}\ L_{\odot}$.
On the other hand, Urquhart et al.\ (2018) estimated a luminosity of $\sim$6.3$\times$10$^{4}\ L_{\odot}$ for this source adopting the parallax distance of 2.9 kpc (Sato et al.\ 2014) and using the spectral energy distribution (SED) obtained from near-infrared to 870 $\mu$m flux by the APEX Telescope Large Area Survey of the Galaxy (ATLASGAL) survey.
It should be noted that the distance of 2.9 kpc is measured for HMSFR G5.88$-$0.39 which is different source apart from G5.900$-$0.340 by $\sim$2.3 arcmin and these are likely to be part of the same molecular cloud complex (Caswell et al.\ 1995, 2010).
In addition, Green et al.\ (2017) derived the distance of these two sources as 2.9 kpc, using Reid et al.\ (2016), which estimated the shapes of the spiral arm by the Bayesian approach.
Thus here we also adopt the same distance of 2.9 kpc for G5.900$-$0.340.
According to Urquhart et al.\ (2018), the mean value of measurement error of luminosity is 42\%.
In addition, the uncertainty due to the assumption of $\beta$ used in SED fitting is effective by a factor of few.
Considering these two errors, the total uncertainty of luminosity is $\sim$100\%.
On the other hand, an uncertainty of the P--L relation arises from possible variations of  protostellar evolution tracks (Inayoshi et al.\ 2013).
According to Inayoshi et al.\ (2013), the P--L relation shown in equation (1) Inayoshi et al.\ (2013) derived from protostellar evolution with spherical accretion model given by Hosokawa and Omukai\  (2009), while accretion via a geometrically thin disk model given by Hosokawa, Yorke, and Omukai\ (2010) give the period of 10 times longer than period in equation (1) in Inayoshi et al.\ (2013). 
Therefore, we cannot completely rule out that the periodic variation of high velocity features is derived from protostellar pulsation.
Thus the protostellar pulsation can be driving the flux variations of the 9.77 and 10.84 km s$^{-1}$ features.
If the pulsationally unstable model is applied for this source,
the protostellar mass and mass accretion rate estimated from the period of $\sim$ 1260 days using the equations (2) and (4) in Inayoshi et al (2013) are $M_{*}=37\ M_{\odot}$ and $\dot{M_{*}}=2\times$10$^{-2}\ M_{\odot}\ {\rm yr}^{-1}$, respectively.

Rotating spiral shock model in a binary system is proposed by Parfenov and Sobolev (2014).
The dust temperature variations are caused by rotation of hot and dense material in the spiral shock wave in the circumbinary disk central gap.
This model does not require either radio or infrared emission variability,
but needs an edge-on protostellar disk which  masers are associated.
According to van der Walt et al.\ (2016), the flux variation pattern in spiral shock model has to show a tail in decaying phase, resulting in non-quiescent phase and difficulty to present an intermittent profile of each flare.
Morgan et al.\ (2021) suggests that the source geometry and orientation are also important factors influencing observed flare profiles.
Therefore, it is important to clarify the spatial distribution of maser features and their flux variation pattern.

Unfortunately, the spatial distribution of this source especially for $V_{\rm LSR} <  \rm{4\ km\ s}^{-1}$ is not observationally revealed and we cannot examine the validity of this model.
Therefore, we cannot judge whether the spiral shock model is appropriate for the flux variation of the 0.66 km s$^{-1}$ feature or not.
High-resolution Very Long Baseline Interferometry (VLBI) observations during the active phase of the 0.66 km s$^{-1}$ feature are required to better understand the periodic variation of this source.

In general, however, pulsation of the central star spreads spherically without directionality, therefore in the pulsation model, all features will vary with the same period.
On the other hand, for the CWB and spiral shock models, the presence or absence of periodic flux variation, and the phase of periodic flux variation may vary from feature to feature, depending on their geometric arrangement.  
Thus the 0.66 km s$^{-1}$ and high-velocity features may be associated with different sources.
VLBI observations are required to obtain the spatial distribution of the 0.66 km s$^{-1}$ and high velocity features to resolve this open question.

\section{Summary}
We present the new periodic 6.7 GHz methanol maser source G5.900$-$0.430, detected from long-term monitoring observations conducted by Ibaraki University.
Periodic flux variabilities was detected at three velocity features among four in this source.
The obtained periods are 130.6 days for low-velocity feature and $\sim$1260 days for two high velocity features.
The period of 1260 days is the longest ever found.
From the intermittent and asymmetric variability profile, the low-velocity feature is likely to be explained by the CWB model or spiral shock model, while the continuous and symmetric variation of the high-velocity features are likely to be caused by the protostellar pulsation.
Consistency with the expected period-luminosity relation also supports the pulsation as the cause of the periodic variation of the high-velocity components.
Simultaneous VLBI observations of the high-velocity and low-velocity components during the active phase of 0.66 km s$^{-1}$ feature will lead to a better understanding of the nature of this source.

\begin{ack}
The authors are grateful to all the staff and students at Ibaraki University who have supported observations of the Ibaraki 6.7 GHz Methanol Maser Monitor (iMet) program.
This work is partially supported by the Inter-university collaborative project ``Japanese VLBI Network (JVN)'' of NAOJ and JSPS KAKENHI Grant Numbers JP24340034, JP21H01120, and JP21H00032 (YY). 
\end{ack}

\end{document}